\documentclass{czjphys}         
\usepackage{euscript,amssymb,epsfig}
\def\RR{{\mathbb R}}

\def\dim{\mbox{\rm dim}}
\def\codim{\mbox{\rm codim}}
\def\nn{\nonumber}
\begin{document}
\title{Third order spectral branch points in Krein space related setups: ${\cal PT}-$symmetric matrix toy
model,\\ MHD $\alpha^2-$dynamo, and extended Squire equation
\footnote{Presented at the 3rd International Workshop
"Pseudo-Hermitian Hamiltonians in Quantum Physics", Istanbul,
Turkey, June 20-22, 2005.}}
\authori{Uwe G\"unther and Frank Stefani}
\addressi{Research Center Rossendorf, Department of
Magnetohydrodynamics,\\
P.O. Box 510119, D-01314 Dresden, Germany}
\authorii{}
\addressii{}
\authoriii{}    \addressiii{}
\authoriv{}     \addressiv{}
\authorv{}      \addressv{}
\authorvi{}     \addressvi{}
\headauthor{Uwe G\"unther et al.} \headtitle{Third order spectral
branch points \ldots} \lastevenhead{Uwe G\"unther et al.: Third
order spectral branch points \ldots}
\refnum{}
\daterec{}    
\issuenumber{9}  \year{2005} \setcounter{page}{1}
\maketitle
\begin{abstract}
The spectra of self-adjoint operators in Krein spaces are known to
possess real sectors as well as sectors of pair-wise complex
conjugate eigenvalues. Transitions from one spectral sector to the
other are a rather generic feature and they usually occur at
exceptional points of square root branching type. For certain
parameter configurations two or more such exceptional points may
happen to coalesce and to form a higher order branch point. We study
the coalescence of two square root branch points semi-analytically
for a ${\cal PT}-$symmetric $4\times 4$ matrix toy model and
illustrate numerically its occurrence in the spectrum of the
$2\times 2$ operator matrix of the magneto-hy\-dro\-dy\-na\-mic
$\alpha^2-$dynamo and of an extended version of the hydrodynamic
Squire equation.
\end{abstract}

\section{Krein space related physical setups and spectral phase transitions}

Some basic spectral properties of the Hamiltonians of ${\cal
PT}-$sym\-me\-tric Quantum Mechanics (PTSQM) \cite{bender-1} can be
easily explained from the fact \cite{japar,LT-1} that these
Hamiltonians are self-adjoint operators in Krein spaces
\cite{azizov,langer-1}
--- Hilbert spaces with an indefinite metric structure. In contrast
to the purely real spectra of self-adjoint operators in "usual"
Hilbert spaces (with positive definite metric structure), the
spectrum of self-adjoint operators in Krein spaces splits into real
sectors and sectors with pair-wise complex conjugate eigenvalues. In
physical terms these two types of sectors are equivalent to phases
of exact ${\cal PT}-$symmetry and spontaneously broken ${\cal
PT}-$symmetry \cite{bender-1}. The reality of the spectrum of a
PTSQM Hamiltonian, e.g., with complex potential $ix^3$, means that
this spectrum is located solely in a real sector and that the
operator is quasi-Hermitian\footnote{Because quasi-Hermitian
operators form only a restricted subclass of self-adjoint operators
in Krein spaces (pseudo-Hermitian operators in the sense of Ref.
\cite{most1}), the question for the reality of the spectrum seems up
to now only partially solved. Apart from the requirement for
existing ${\cal PT}-$symmetry of the differential expression of the
operator, a subtle interplay between the operator domain and some,
in general not yet sufficiently clearly identified, additional
structural aspects of its differential expression seems to be
responsible for the quasi-Hermiticity of the operator (exact ${\cal
PT}-$symmetry of the corresponding PTSQM setup).} in the sense of
Ref. \cite{geyer}. In general, the spectrum of a self-adjoint
operator in a Krein space is spreading over both sectors.

Considering the spectrum as a Riemann surface depending on model
parameters (not necessarily moduli) from a space ${\cal M}$, the
boundaries between real and complex spectral sectors can be
described as algebraic variety ((multi-component) hypersurface)
$\Upsilon$ in ${\cal M}$, $\Upsilon \subset {\cal M}$, where the
Riemann surface has exceptional points of branching type
\cite{kato,baumg,heiss-1,heiss-3,rotter,berry-2,heiss-2}. The most
generic varieties of this type are those of codimension one, $\codim
(\Upsilon)=\dim ({\cal M})-\dim (\Upsilon)=1$, what can be easily
read off, e.g., from a pseudo-Hermitian $2\times 2-$matrix model
\cite{MZ-sep,most-npb-1,BBJ-1,GSG-cz2}
\be\label{1} H=\left(\begin{array}{cc}
               x+y & w \\
                 -w^* & x-y \\
               \end{array}\right),\quad H\psi=\lambda\psi
\ee
where $\Upsilon$ is simply a double cone $y^2=|w|^2\subset {\cal
M}\ni (y,\Re w,\Im w)$. Such codimension-one varieties correspond to
a pairwise real-to-complex transition of two eigenvalues what in
Riemann surface terms is equivalent to a square root branching of
two of its sheets (two spectral branches). At the same time the two
(geometric) eigenvectors coalesce into a single (geometric)
eigenvector, and an additional associated vector (algebraic
eigenvector) appears. Instead of {\it two} eigenvalues of geometric
and algebraic multiplicity one (diagonal block), a {\it single}
eigenvalue of geometric multiplicity one, $(m(\lambda)=1)$, and
algebraic multiplicity two $(n(\lambda)=2)$, ($2\times 2-$Jordan
block) forms \cite{GSG-cz2}. When the parameters on this
codimension-one variety are further tuned to $y=\Re w=\Im w=0$
\cite{GSG-cz2} one arrives at a diabolic point
\cite{berry-3,berry-1} where the {\it single} eigenvalue gets
geometric and algebraic multiplicity two ($m(\lambda)=n(\lambda)=2$,
diagonal $2\times 2-$ block). In a (more) general setting such
configurations correspond to codimension-three varieties
\cite{GSG-cz2,berry-1}.

Apart from these generic real-to-complex transitions on
codimension-one varieties, there may occur higher order
intersections of more than two Riemann sheets simultaneously
--- on varieties $\Upsilon$ of higher codimension, $\codim (\Upsilon)\ge 2$, and with larger
Jordan blocks in the spectral decomposition
\cite{baumg,mondragon-1,maily-1}.

Below we present explicit examples for third order intersections in
three Krein space related physical models. First, we sketch a few
aspects of such intersections semi-analytically (algebraically) for
a maximally simplified ${\cal PT}-$symmetric $4\times 4 -$matrix toy
model. Afterwards, we show numerically that such intersections also
occur in the operator spectra of the spherically symmetric
$\alpha^2-$dy\-na\-mo of magnetohydrodynamics (MHD)
\cite{krause,GS-1} and of the recently analyzed ${\cal
PT}-$sym\-me\-tric interpolation model of Ref. \cite{GSZ-squire}
(which can be understood as a ${\cal PT}-$symme\-tri\-cal\-ly
extended, rescaled and Wick-rotated version of the Squire equation
of hydrodynamics).

\section{Spectral triple points in a pseudo-Hermitian $4\times 4-$matrix toy model}

Similar to the two different multiplicity contents (Jordan
structures) of two-fold degenerate eigenvalues (see, e.g., also
\cite{korsch}), one has to distinguish the following three types of
Jordan structures/geometric multiplicities $m(\lambda)$
corresponding to a spectral triple point:
\begin{itemize}
\item type I: $m(\lambda)=1$, $n(\lambda)=3$; $3\times 3-$Jordan
block; e.g., coalescence of two square-root branch points connected
by a purely real spectral segment (see Eq. (\ref{m-13}) and Figs.
\ref{fig1} - \ref{fig3} below),
\item type II: $m(\lambda)=2$, $n(\lambda)=3$; one $2\times 2-$Jordan
block $+$ one simple eigenvalue; e.g., a square-root branch point
accidentally coincides with a simple eigenvalue,
\item type III: $m(\lambda)=3$, $n(\lambda)=3$; diagonal matrix $\lambda I_3$; e.g., three accidentally
coinciding eigenvalues (generalized diabolic point).
\end{itemize}
Subsequently, we concentrate on the physically most interesting case
of type I triple points which correspond to pair-wise coalescing
square-root branch points.

As maximally simplified toy model we choose a ${\cal PT}-$symmetric
(pseudo-Her\-mi\-tian) $4\times 4-$matrix setup in a diagonal
representation of the parity (involution) operator ${\cal P}$
\be\label{m-1}
H={\cal P}H^\dagger{\cal P}=\left(\begin{array}{cc}
                                             H_{++} & H_{+-} \\
                                             H_{-+} & H_{--} \\
                                           \end{array}\right),\quad {\cal P}=\left(\begin{array}{cc}
                                             I_2 & 0 \\
                                             0 & -I_2 \\
                                           \end{array}\right)\, ,
\ee
where the $2\times 2-$blocks satisfy the conditions
$ H_{\pm\pm}=H_{\pm\pm}^\dagger,\quad H_{+-}=-H_{-+}^\dagger\, $.
In order to keep the demonstration as simple as possible, it
suffices to consider a model based on two real (truncated)
elementary pseudo-Hermitian $2\times 2-$matrices
\be \label{m-2}
H_{1,2}=\left(\begin{array}{cc}
         x_{1,2}+y_{1,2} & w_{1,2} \\
                -w_{1,2} & x_{1,2}-y_{1,2} \\
              \end{array}\right), \qquad x_{1,2},y_{1,2},w_{1,2}\in \RR
\ee
as subsystems, which we embed ${\cal PT}-$symmetrically into the
also highly reduced\footnote{Nine of the 16 effective free real
parameters of the pseudo-Hermitian $4\times 4-$matrix $H$ are set to
zero.} and real pseudo-Hermitian $4\times 4-$matrix $H$ with block
structure (\ref{m-1})\be \label{m-3} H_1,H_2\hookrightarrow
H=\left(\begin{array}{cccc}
                           x_1+y_1 & 0 & w_1 & z \\
                           0 & x_2+y_2 & 0 & w_2 \\
                           -w_1 & 0 & x_1-y_1 & 0 \\
                           -z & -w_2 & 0 & x_2-y_2 \\
                         \end{array}\right)\, .
\ee
The interaction between the subsystems $H_1$ and $H_2$ is controlled
by the coupling parameter $z$. For vanishing coupling, $z=0$, the
characteristic equation
$ \Delta(\lambda)=\det(H-\lambda I_4)=0$
factors as $\Delta(\lambda)=\det(H_1-\lambda I_2)\det(H_2-\lambda
I_2)=0$, and the four eigenvalues of the matrix $H$ are defined by
the eigenvalues of $H_1$ and $H_2$
\be \label{m-5}\lambda_{1,\pm}=x_1 \pm \sqrt{y_1^2-w_1^2},\quad \lambda_{2,\pm}=x_2 \pm
\sqrt{y_2^2-w_2^2}\, .
\ee
The corresponding two square root branch points $\lambda_1=x_1, \
\lambda_2=x_2$ with $ \Delta(\lambda)=0, \quad
\partial_\lambda\Delta(\lambda)=0 $
are located on the two (reduced) double cones\footnote{The two
diabolic points \cite{berry-3,berry-1} (with eigenvalues of
geometric and algebraic multiplicity two) of the subsystems $H_1$,
$H_2$ are located at $y_1=w_1=0$ and $y_2=w_2=0$, respectively.}
(crossed lines) defined by
\be\label{m-6} y_1^2=w_1^2,\qquad y_2^2=w_2^2.
\ee
For non-vanishing interaction $z\neq 0$, the eigenvalues of $H$ are
given as roots of the quartic equation
\be\label{m-7}
\Delta(\lambda)=\lambda^4+a_3\lambda^3+a_2\lambda^2+a_1\lambda+a_0=0
\ee
with coefficients\footnote{The reason for restricting to the
maximally simplified (truncated) $4\times 4-$matrix setup
(\ref{m-3}) was in keeping these coefficients $a_k,\ k=0,\ldots, 3$
in a form sufficiently simple for quick inspection.}
\bea\label{m-8}
a_3&=&-2(x_1+x_2)\nn\\
a_2&=&-(y_1^2-w_1^2)-(y_2^2-w_2^2)+(x_1+x_2)^2+2x_1x_2+z^2\nn\\
a_1&=&2\left[x_1(y_2^2-w_2^2-x_2^2)+x_2(y_1^2-w_1^2-x_1^2)\right]-z^2(x_1-y_1+x_2+y_2)\nn\\
a_0&=&(y_1^2-w_1^2-x_1^2)(y_2^2-w_2^2-x_2^2)+z^2(x_1-y_1)(x_2+y_2)\,
.
\eea
A smooth branch point coalescence can be easily arranged by ${\cal
PT}-$symmetrically blowing up a triple root
\be \label{m-9} \lambda_1=\lambda_2=\lambda_3=:\beta_{(c)}\neq
\lambda_{4(c)}
\ee
("inflection point" configuration $\Delta(\lambda)=\partial_\lambda
\Delta(\lambda)=\partial^2_\lambda \Delta(\lambda)=0$) of the
quartic equation (\ref{m-7}).
\bfg[htb]                     
\bc                         
\epsfig{file=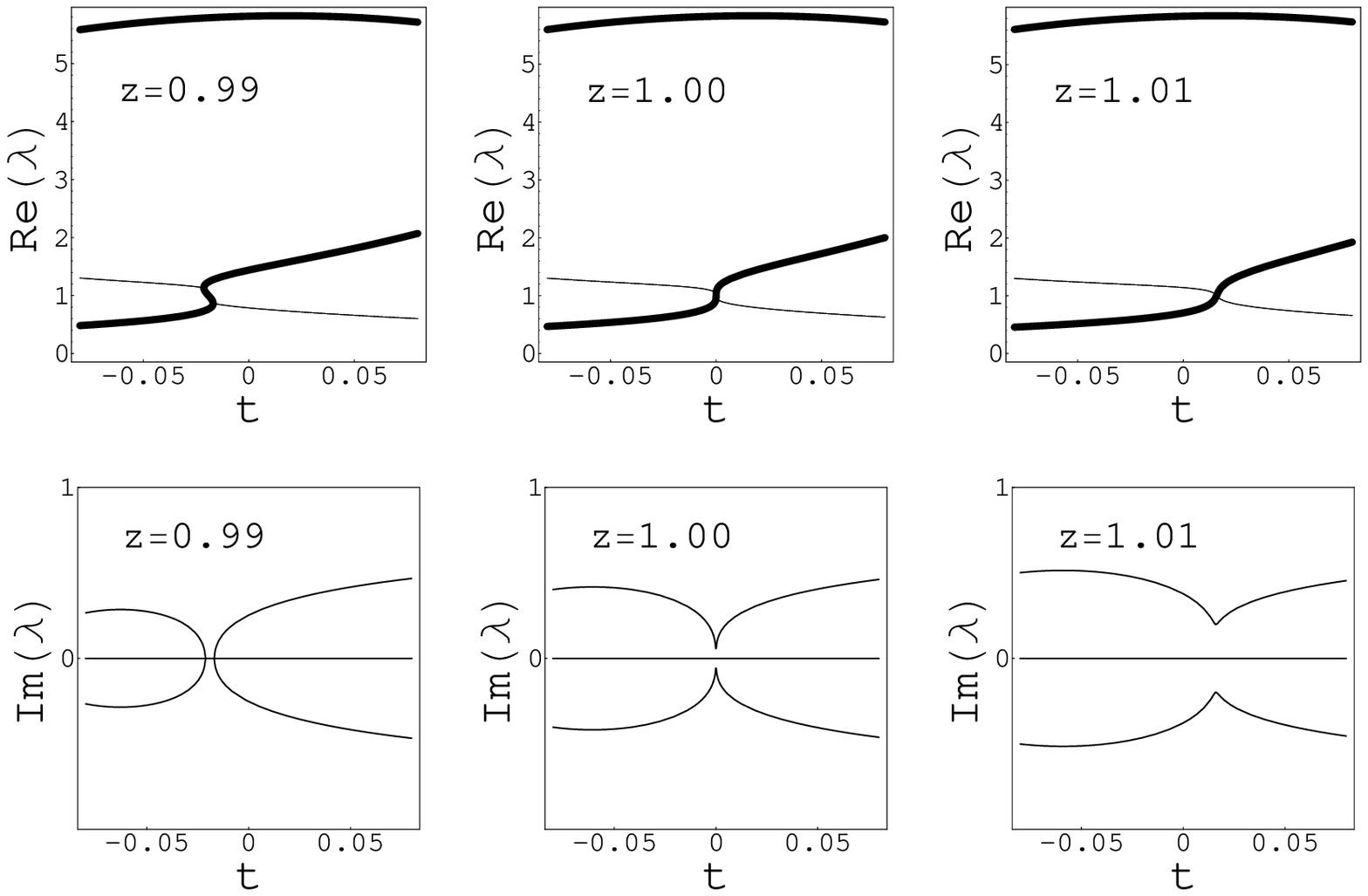,angle=0, width=0.98\textwidth}
\ec                         
\vspace{-8mm}\caption{The four solutions
$\lambda_1,\ldots,\lambda_4$ of the quartic equation
$\Delta(\lambda)=0$ (purely real branches highlighted fat) for
different subsystem couplings $z$. For increasing $z$ two of the
exceptional points coalesce and disappear [at the inflection point
of the inverted purely real branch $t(\Re \lambda_1)$: \
$\partial^2_{\Re \lambda_1}t(\Re \lambda_1)=0$]. Simultaneously, the
originally existing real spectral segment between the exceptional
points disappears and the two complex valued sectors merge into a
single one.\label{fig1}}
\efg 
For our illustration purpose, it suffices to explicitly derive a
suitable parametrization for an arbitrarily chosen triple
root\footnote{A detailed discriminant-based \cite{discri} case
analysis and complete classification of possible intersection
scenarios for the four roots $\lambda_k,\ k=1,\ldots, 4$ will be
presented elsewhere.} satisfying the additional branch point
conditions (\ref{m-6}) for the subsystems. Using these conditions
together with (\ref{m-8}), (\ref{m-9}) in Newton's identities
(Vieta's formulas) \cite{waerden} \bea \label{m-10}
\lambda_1+\lambda_2+\lambda_3+\lambda_4&=&-a_3\nn\\
\lambda_1\lambda_2+\lambda_1\lambda_3+\lambda_1\lambda_4
+\lambda_2\lambda_3+\lambda_2\lambda_4+\lambda_3\lambda_4&=&a_2\nn\\
\lambda_1\lambda_2\lambda_3+\lambda_2\lambda_3\lambda_4+\lambda_1\lambda_2\lambda_4+\lambda_1\lambda_3\lambda_4&=&-a_1\nn\\
\lambda_1\lambda_2\lambda_3\lambda_4&=&a_0
\eea
and setting for convenience $\beta_{(c)}=z_{(c)}=1,\quad x_{1(c)}=0$
we easily obtain the remaining root $\lambda_{4(c)}$ and the three
parameters $x_{2(c)},y_{1,2(c)}$  as
\bea\label{m-12}\lambda_{4(c)\epsilon}&=&3+ 2^{3/2}\epsilon>0,\quad
\epsilon:=\pm 1\, ,\nn\\
x_{2(c)\epsilon}&=&\frac 12 (\lambda_{4(c)\epsilon}+3)=3+
2^{1/2}\epsilon\, ,\nn\\
y_{1(c)\epsilon,\delta}&=&\frac 12\left[-(3\lambda_{4(c)\epsilon}+1)
+\delta \sqrt{9\lambda^2_{4(c)\epsilon}+2\lambda_{4(c)\epsilon}+1}\right],\quad \delta:=\pm 1\, ,\nn\\
y_{2(c)\epsilon,\delta}&=&\lambda_{4(c)\epsilon}-1+\frac\delta 2
\sqrt{9\lambda^2_{4(c)\epsilon}+2\lambda_{4(c)\epsilon}+1}\, .
\eea
A computer algebraic test shows that the Jordan normal form of $H$
for this triple root configuration reads \be\label{m-13}
H_{(c)}=SH_{(c)J}S^{-1},\qquad H_{(c)J}=\left(\begin{array}{cccc}
                                          1 & 1 & 0 & 0 \\
                                          0 & 1 & 1 & 0 \\
                                          0 & 0 & 1 & 0 \\
                                          0 & 0 & 0 & \lambda_{4(c)\epsilon} \\
                                        \end{array}\right),
\ee
so that the eigenvalue $\beta_{(c)}=1$ is indeed a type I triple
point with  geometric multiplicity one and algebraic multiplicity
three. For the $2D-$illustration of the coalescing branch points in
Fig. \ref{fig1} we have chosen a one-parameter matrix
parametrization of the type $ H(t)=H_{(c)}+h(t),\ \ h(t\to 0)\to 0 $
providing the blowing-up of the triple root for $t\neq 0$ with
\bea\label{m-14}
& x_1=x_{1(c)}\, ,\qquad & x_2=x_{2(c)}+t\, ,\nn\\
& y_1=y_{1(c)++}+2t^2\, \qquad & y_2=y_{2(c)++}-t^3\, ,\nn\\
& w_1=y_{1(c)++}-t\, \qquad & w_2=y_{2(c)++}+3t^2\, .
\eea
A characteristic feature of the present coalescing branch point
setup consists in the merging of two originally separated complex
spectral sectors (of broken ${\cal PT}-$sym\-me\-try) into a larger
single sector and the simultaneous disappearance of the {\it real
segment} between the two square-root branch points. This is in
obvious contrast to coalescing branch point setups in simpler
pseudo-Hermitian $2\times 2-$matrix models, where the branch points
are connected by {\it two real or complex branches} which disappear
upon branch point coalescence.

\section{Third order spectral transitions of the $\alpha^2-$dynamo and the extended Squire setup}

Based on the information of the previous section, it is easy to
identify configurations with coalescing branch points in the spectra
of the MHD $\alpha^2-$dynamo operator and of the operator of the
extended Squire setup.

In case of the operator \cite{GSG-cz2,GS-1,GSZ-squire} of the
spherically symmetric $\alpha^2-$dynamo \cite{krause}
\be \hat
H_l[\alpha]=\left(\begin{array}{cc}-Q[1] & \alpha\\ Q[\alpha] &
-Q[1]
 \end{array}\right),\quad Q[\alpha]:=-(\partial_r +1/r)\alpha(r)
(\partial_r +1/r) + \alpha(r) \frac{l(l+1)}{r^2}\label{d1}
\ee
with $\alpha-$profile\footnote{The rather special numerical
coefficients in the $\alpha-$profile are adopted from the recently
studied field-reversal scenario for $\alpha^2-$dynamos
\cite{reversal}.}
\bea \label{d2} \alpha(r)&=&C \left[- (21.465+2.467 \zeta) +
(426.412+ 167.928 \zeta
) r^2\right.\nn\\
&&\left. - (806.729+ 436.289 \zeta ) r^3 + (392.276+272.991 \zeta
) r^4 \right]
\eea
\bfg[htb]                     
\bc                         
\epsfig{file=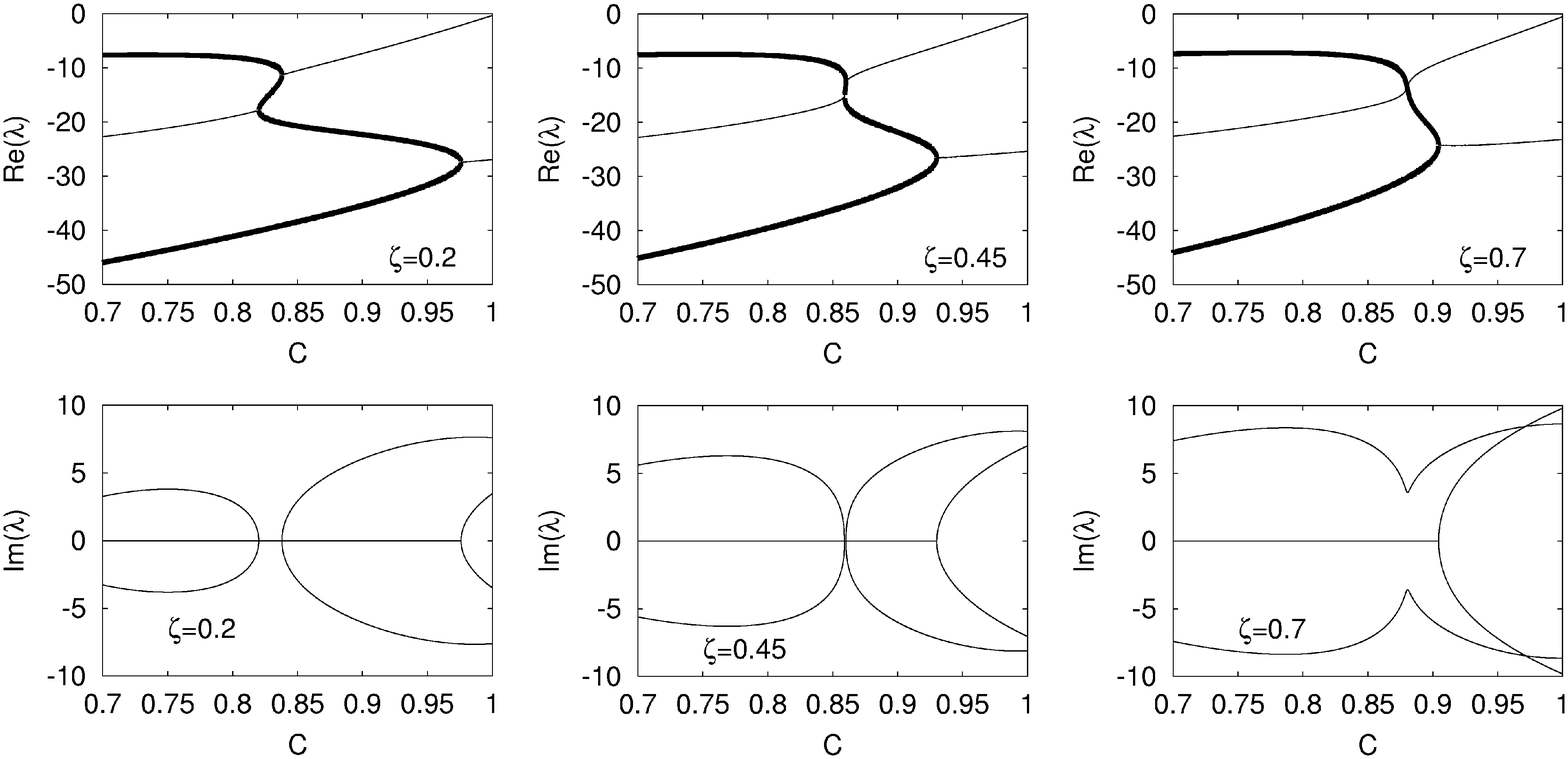,angle=0, width=0.98\textwidth}
\ec                         
\vspace{-2mm} \caption{Coalescing branch points in the spectrum of
an $\alpha^2-$dynamo with $\alpha-$profile (\ref{d1}) depending on a
warp parameter $\zeta$. \label{fig2}}
\efg                        
and physical (realistic) boundary conditions
\cite{GSG-cz2,krause,GS-1,GSZ-squire} a triple point transition in
the decay-mode sector $(\Re \lambda <0)$ is depicted in Fig.
\ref{fig2}. Up to now it is still an open question whether such a
transition in the sector of growing modes $(\Re \lambda >0)$ will
have any physical significance, e.g., in dynamo experiments
\cite{riga,karlsruhe}.

A different situation occurs in the case of the extended Squire
setup of Ref. \cite{GSZ-squire} \be\label{d3} \left[ -\partial_y^2
+g\,y^2(iy)^\nu \right] \psi(y) = E\, \psi(y),\qquad \psi(y=\pm b)=0
\ee
\bfg[htb]                     
\bc                         
\epsfig{file=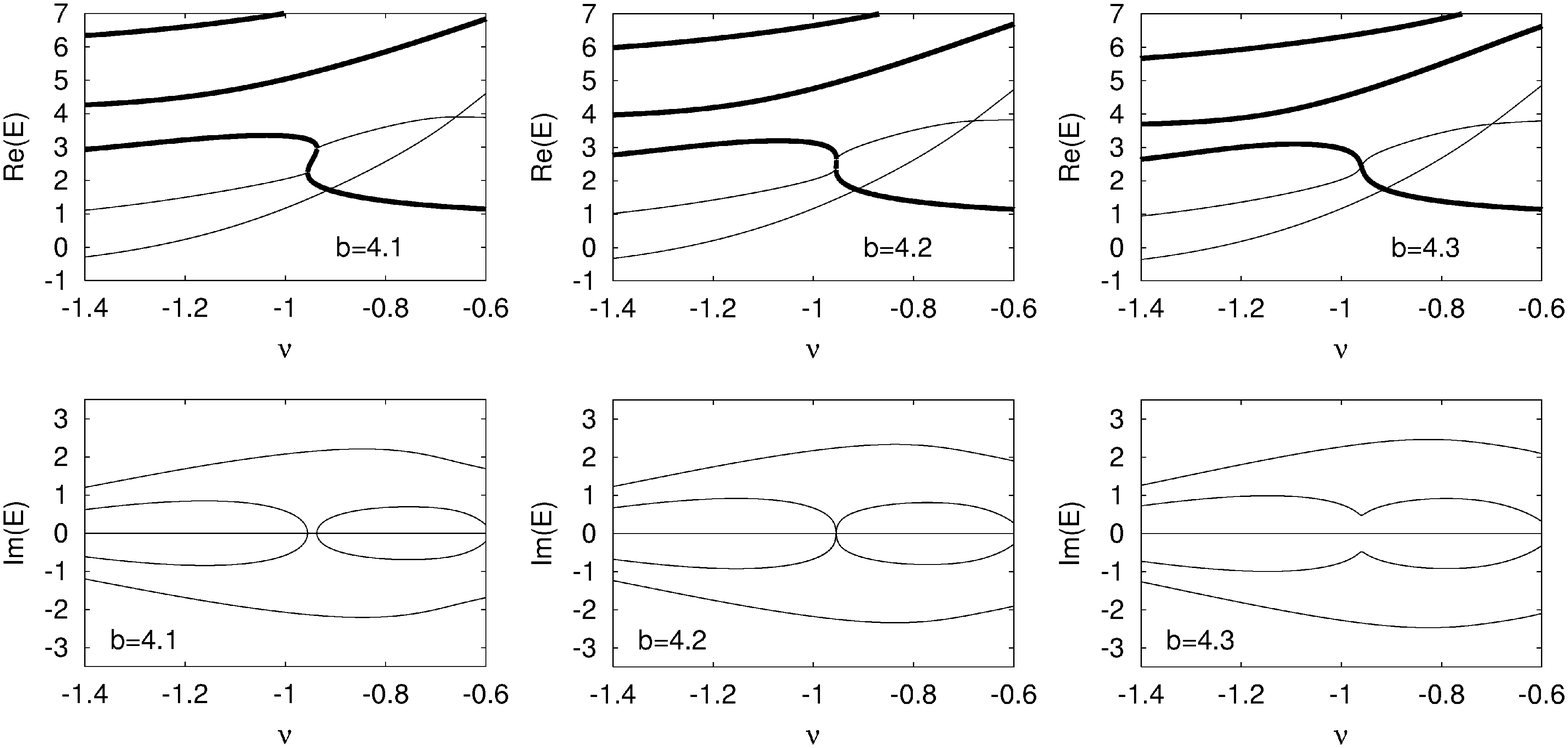,angle=0, width=0.98\textwidth}
\ec                         
\vspace{-2mm} \caption{Coalescing branch points in the spectrum of
the extended Squire setup (\ref{d3}) with unit coupling, $g=1$, in
dependence of the cut-off/box length $b$. \label{fig3}} \efg which
corresponds to the Bender-Boettcher problem \cite{bender-1} over a
finite interval $[-b,b]$. There a coalescence of two square-root
branch points occurs close to the real-to-complex transition point
in the associated Herbst-box model \cite{GSZ-squire} (Herbst model
\cite{Herbst-1} over a finite interval) located at the low-energy
end of a web-like branch structure. An example for the corresponding
triple point transition (which supplements the qualitative and
graphical analyzes of Ref. \cite{GSZ-squire}) is depicted in Fig.
\ref{fig3}.

\bigskip
{\small This work has been supported by DFG grant GE 682/12-2.}

\end{document}